\documentclass[twocolumn,english,aps,pra,superscriptaddress,showpacs,tightenlines]{revtex4-1}
\usepackage{amsmath}
\usepackage{graphicx}
\usepackage{amssymb}

\begin{document}

\title{Single-photon router:  Coherent Control of multi-channel scattering for
single-photons with quantum interferences}

\author{Jing \surname{Lu}}
\affiliation{Key Laboratory of Low-Dimensional Quantum Structures and Quantum
Control of Ministry of Education, and Department of Physics, Hunan
Normal University, Changsha 410081, China}
\affiliation{CEMS, RIKEN, Saitama, 351-0198, Japan}
\author{Lan \surname{Zhou}}
\thanks{Corresponding author}
\email{zzhoulan@gmail.com}
\affiliation{Key Laboratory of Low-Dimensional Quantum Structures and Quantum
Control of Ministry of Education, and Department of Physics, Hunan
Normal University, Changsha 410081, China}
\affiliation{Beijing Computational Science Research Center, Beijing 100084, China}
\author{Le-Man \surname{Kuang}}
\affiliation{Key Laboratory of Low-Dimensional Quantum Structures and Quantum
Control of Ministry of Education, and Department of Physics, Hunan
Normal University, Changsha 410081, China}
\author{Franco Nori}
\affiliation{CEMS, RIKEN, Saitama, 351-0198, Japan}
\affiliation{Physics Department, The University of Michigan, Ann Arbor, MI 48109-1040, USA.}

\begin{abstract}
We propose a single-photon router using a single atom with an inversion center
coupled to quantum multichannels made of coupled-resonator waveguides.
We show that the spontaneous emission of the atom can direct
single photons from one quantum channel into another. The on-demand classical
field perfectly switches-off the single-photon routing due to the quantum
interference in the atomic amplitudes of optical transitions. Total reflections
in the incident channel are due to the photonic bound state in the continuum.
Two virtual channels, named as scatter-free and controllable channels, are found,
which are coherent superpositions of quantum channels. Any incident photon in
the scatter-free channel is totally transmitted. The propagating states of the
controllable channel are orthogonal to those of the scatter-free channel. Single
photons in the controllable channel can be perfectly reflected or transmitted by the atom.
\end{abstract}

\maketitle

\narrowtext

\section{Introduction}

A quantum network~\cite{nwork} consists of quantum channels and nodes, which
are provided by waveguides and quantum emitters, respectively. The flying
qubits in quantum channels serve to distribute quantum information. The
static qubits in local nodes generate, process, and route quantum
information. Photons are ideal carriers in quantum channels because they are
fast, robust, and readily available. Although it is easy to control photons
in linear optical systems~\cite{Halllos,Lemrlos1,Duanlos2,Duanlos},
waveguides are more promising as the technology proceeds to smaller on chip
structures. Currently, considerable attention has been paid to photon
transport in 1D waveguides with a quantum emitter both in theory~\cite%
{FanQdev,zhouQdev,Supcavity,InMZ,attenuator,
MTcheng,zhqrouter,wgQEDqi,Fan2pho,Longo,Alexanian,TShiSun,mbbound,
Roybound,Fan2sm,Yanc2ph,HOMeff} and experiments~\cite%
{g-plasmon,TAPRL102,pcfiber,dnwire,
CircuitQED,GaAspnwire1,GaAspnwire2,transmon}. The waveguide confines photons
in low dimension, and has a dispersion relation different from photons in
free space, giving rise to new physical phenomena, e.g. the total reflection
in Ref.~\cite{FanQdev,zhouQdev}. The coupled-resonator waveguide (CRW)~\cite%
{Plenio} is an important waveguide for studying waveguide-QED due to the
following advantages: (1) strong coupling between light and matter; (2)
wider bandwidths; (3) scalability (e.g., ultrahigh-$Q$ coupled nanocavity
arrays with $N>100$ has been realized in photonic crystals~\cite{scaleCRW});
(4) easy addressability; (5) ability to simulate the behavior of single
particles in the short- and long-wavelength regimes. The discreteness of CRW
offers a rich variety of properties and possibilities that do not exist in
the bulk, e.g., the bound states outside the band~\cite{qzenos}. Currently,
many photonic devices (see, e.g.,~\cite%
{zhouQdev,Supcavity,InMZ,attenuator,zhqrouter} and references therein) based
on using single atoms in a 1D CRW have been proposed. Most of the
theoretical work focuses on controlling photons in one continuum of
propagating states. Recently, a cyclic three-level system embedded in
multiple quantum channels formed with 1D CRWs has been proposed as a quantum
router~\cite{zhqrouter}, where single photons are routed from one continuum
to another by the cyclic system with the help of a classical control field.
However, the experimental realization of this proposal faces the challenge
that cyclic transitions are forbidden for natural atoms as these possess an
inversion center~\cite{cyclicA}. One should use either chiral systems~\cite%
{Ylicyc1} or atoms whose symmetry is broken artificially~\cite%
{Liucycl,Ylicyc2}. These bring complexities to a possible realization. A
router using natural atoms will certainly contribute to the studies on
quantum networks and routers, and facilitate its experimental realization.

In this work, we propose a single-photon routing scheme using systems with
an inversion center. The two continuum of propagating states are constructed
by two 1D CRWs. To control the transfer of propagating states from one
continuum to the other, we explore quantum coherence and interference
effects, such as the electromagnetically induced transparency (EIT) in a
system of a three-level $\Lambda$ atom embedded inside the two 1D CRWs. Here
the $\Lambda$ atom plays the role of a quantum node for routing. One
transition of the $\Lambda$ atom is coupled to the photonic modes of the two
CRWs. The other transition is driven by a classical field. Different from
the routers~\cite{Creffield,Zueco} based on designing a time-dependent
classical field acting on an large area of the considered system, here, the
classical field is applied to individually address the atom. The scattering
process is studied when a single photon is incident from one CRW. We find
that the quantum node indeed works as a multi-channel quantum router, and
the classical field selects the channel where single photons are directed
or transferred to. The multi-channel effect is taken into account by studying
the single-photon scattering process with waves incident from two CRWs. A
controllable channel and a scattering-free channel~\cite{wgQEDqi} are found
when both CRWs are identical.

This paper is organized as follows: In Sec.~\ref{Sec:2}, we introduce our
model, which consists of a three-level $\Lambda$ atom embedded in two CRWs.
In Sec.~\ref{Sec:3}, we employ the discrete-coordinate scattering approach
to study the scattering process of single photons and give the expressions
for scattering amplitudes. We discuss the function of the three-level atom
in Sec.~\ref{Sec:4} using the eigenstates for the total system. Here, two
different band configurations of the two CRWs are studied, and the
underlying physics for controlling single-photon routing are discussed. In
Sec.~\ref{Sec:5}, we study the possibility for a three-level atom to act as
a perfect mirror or a transparent medium for single photons with waves
incident from two CRWs. Finally, we conclude with a brief summary of the
results with accompanying discussions.


\section{\label{Sec:2}model setup}


As shown in Fig. 1, our hybrid system consists of two 1D CRWs and a
three-level system. The cavity modes of the two 1D CRWs are described by the
annihilation operators $a_{j}$ and $b_{j}$, respectively, and subscript $%
j=-\infty,\cdots,+\infty$. The atom located at $j=0$ is characterized by the
ground state $|g\rangle$, one intermediate state $|s\rangle$, and an excited
state $|e\rangle$. The transition $|g\rangle\leftrightarrow|e\rangle$ is
dipole-coupled to the cavity modes $a_{0}$ and $b_{0}$, with coupling
strengths $g_a$ and $g_b$, respectively. Obviously, the cavity-driven
transition builds a bridge between these two CRWs. The classical field with
frequency $\nu$ drives the atomic transition $|s\rangle\leftrightarrow|e\rangle$
with Rabi frequency $\Omega$. The transition between the ground and
intermediate states are forbidden.
\begin{figure}[tbp]
\includegraphics[bb=39bp 461bp 439bp 770bp,width=8cm]{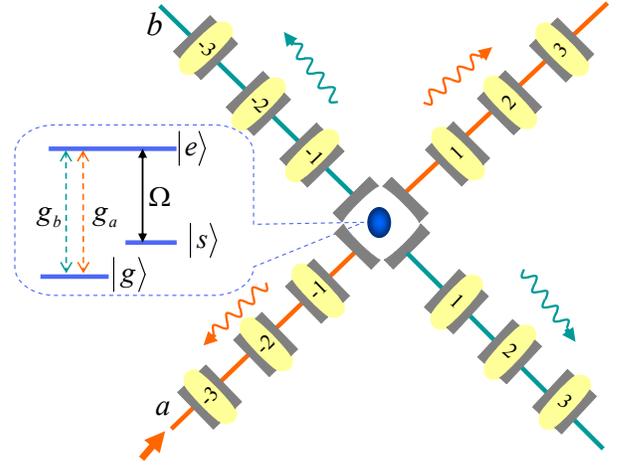}
\caption{(Color online) Schematic routing of single photons in two channels
made of two CRWs. The three-level atom characterized by $\left|g\right%
\rangle $, $\left|e\right\rangle $, and $\left|s\right\rangle $ is placed at
the cross point $j=0$. CRW-$a$ (-$b$) couples to the atom through the
transition $\left|g\right\rangle \leftrightarrow\left|e\right\rangle $ with
strength $g_{a}$ ($g_{b}$) and a classical field with Rabi frequency $\Omega$
is applied to drive the $\left|e\right\rangle
\leftrightarrow\left|s\right\rangle $ transition. An incoming wave from the
left side of the CRW-$a$ will be reflected, transmitted, or transferred to
the CRW-$b $.}
\label{fig:1}
\end{figure}

Once a photon is inside one cavity of the CRW, it propagates along the CRW
and is also scattered by the atoms. The total Hamiltonian of this hybrid
system $H=H_{C}+H_{A}^{\prime}+H_{CA}$ contains three parts: The Hamiltonian
$H_{C}$ describes the two CRWS, $H_{A}^{\prime}$ is free Hamiltonian of the $%
\Lambda $-type three-level atom, and $H_{CA}$ describes the interactions
between the cavity modes, the classical field and the atom. The two CRWs are
modeled as two independent linear chains of sites with a nearest-neighbor
interaction, which are described by the Hamiltonian
\begin{align}
H_{C}=\sum_{j}\left[ \omega _{a}a_{j}^{\dagger }a_{j}-\xi _{a}\left(
a_{j}^{\dagger }a_{j+1}+\mathrm{h.c.}\right) \right]  \notag \\
+\sum_{j}\left[ \omega _{b}b_{j}^{\dagger }b_{j}-\xi _{b}\left(
b_{j}^{\dagger }b_{j+1}+\mathrm{h.c.}\right) \right]  \label{2model-01}
\end{align}%
For simplicity, we assume that all resonators in the CRW $a$ $(b)$ have the
same frequency $\omega _{a}$ $(\omega _{b})$ and the hopping energies $\xi
_{a}$ $(\xi _{b})$ between any two nearest-neighbor cavities in the CRW $a$ $%
(b)$ are the same. By introducing the Fourier transform $d_k=\frac{1}{\sqrt{%
2\pi}}d_j e^{ik_dj},d=a,b$, we see that each bare CRW supports plane waves
with the dispersion relation
\begin{subequations}
\begin{align}
E_{k}^{[a]}=\omega _{a}-2\xi _{a}\cos k_a\text{,} \\
E_{k}^{[b]}=\omega _{b}-2\xi _{b}\cos k_b\text{,}  \label{2model-02}
\end{align}
which indicates that each CRW possesses an energy band with bandwidth $4\xi
_{a}$ and $4\xi _{a}$, respectively. Consequently, two continuums are
formed. In Fig.~\ref{fig:1}, all the resonators connected by the red (blue)
line form the photonic channel $a$ ($b$), which is referred to as CRW-a (b)
hereafter in this paper. We note that the central cavity with the atom comprises
two cavities (see Fig.1), one lies on the red line, which is described by
the bosonic destruction operator $a_0$, the other lies on the blue line,
which is described by the bosonic destruction operator $b_0$. The Hamiltonian
for the free $\Lambda $-type three-level atom reads
\end{subequations}
\begin{equation}
H_{A}^{\prime }=\omega _{E}|e\rangle \langle e|+\omega _{S}^{\prime
}|s\rangle \langle s|,  \label{2model-03}
\end{equation}%
where we have chosen the energy of the ground state $\left\vert
g\right\rangle $ as the energy reference. The interaction Hamiltonian
\begin{equation}
H_{CA}=\left\vert e\right\rangle \left\langle g\right\vert \left(
g_{a}a_{0}+g_{b}b_{0}\right) +\Omega \left\vert e\right\rangle \left\langle
s\right\vert e^{-i\nu t}+h.c.  \label{2model-04}
\end{equation}%
is written under the rotating wave approximation. We note here that the classical
field only acts on the atom. To remove the time-dependent factor of the Hamiltonian,
we rewrite the Hamiltonian in a rotating frame of reference, which is defined by
the unitary transformation $U=\exp{(i\nu t\left\vert s\right\rangle \left\langle s\right\vert )}$.
The Hamiltonian $H_{R}\equiv U^{\dag }HU-iU^{\dag }\partial _{t}U$ in this
rotating frame still consists of three parts. The part for the CRWs remain
the same. The free $\Lambda $-type atom transforms to
\begin{equation}
H_{A}=\omega _{E}|e\rangle \langle e|+\omega _{S}|s\rangle \langle s|,
\label{2model-05}
\end{equation}%
where $\omega _{S}=\omega _{S}^{\prime }+\nu $ is the frequency sum of the
intermediate state and the classical light field. The time-dependent
interaction Hamiltonian becomes
\begin{equation}
H_{I}=\left\vert e\right\rangle \left\langle g\right\vert \left(
g_{a}a_{0}+g_{b}b_{0}\right) +\Omega \left\vert e\right\rangle \left\langle
s\right\vert +h.c.,  \label{2model-06}
\end{equation}
which is independent of time. Hereafter, we study the single-photon scattering
in this rotating frame. We note that when the classical field is absent, i.e.,
$\Omega=0$, this system becomes a two-level atom embedded in two CRWs.


\section{\label{Sec:3} Coherent scattering of single photons}


It can be found that the operator
\begin{equation}
N=\sum_{j}(a_{j}^{\dag }a_{j}+b_{j}^{\dag }b_{j})+\left\vert e\right\rangle
\left\langle e\right\vert +\left\vert s\right\rangle \left\langle
s\right\vert   \label{3cs-01}
\end{equation}%
commutes with the Hamiltonian $H_{R}$. Since the number of quanta is
conserved in this hybrid system, three mutually exclusive possibilities are
involved in the one-quantum subspace: (1) The particle freely propagates in
the two CRWs; (2) The particle is absorbed by the atom, consequently, the
atom is populated in its excited state; (3) The classical field stimulates
the atom into its intermediate state. This implies the following stationary
eigenstate%
\begin{eqnarray}
\left\vert E\right\rangle  &=&\sum_{j}\alpha \left( j\right) a_{j}^{\dag
}\left\vert g0\right\rangle +\sum_{j}\beta \left( j\right) b_{j}^{\dag
}\left\vert g0\right\rangle   \label{3cs-02} \\
&&+u_{e}\left\vert e0\right\rangle +u_{s}\left\vert s0\right\rangle ,  \notag
\end{eqnarray}%
where $\left\vert 0\right\rangle $ is the vacuum state of the two CRWs.
Here, $\alpha (j)$ and $\beta (j)$ are the probability amplitudes of
single-photon states in the $j$th cavity of the CRW-$a$ and CRW-$b$
respectively. Also, $u_{e}$ and $u_{s}$ are the probability amplitudes of
the three-level system in its excited and intermediate states, respectively.

The eigenequation gives rise to a series of coupled stationary equations for
all amplitudes
\begin{align*}
\left( E-\omega _{E}\right) u_{e}& =\Omega u_{s}+g_{a}\alpha \left( 0\right)
+g_{b}\beta \left( 0\right), \\
\left( E-\omega _{S}\right) u_{s}& =\Omega ^{\ast }u_{e}, \\
\left( E-\omega _{a}\right) \alpha \left( j\right) & =-\xi _{a}\left[ \alpha
\left( j-1\right) +\alpha \left( j+1\right) \right] +\delta _{j0}g_{a}u_{e},
\\
\left( E-\omega _{b}\right) \beta \left( j\right) & =-\xi _{b}\left[ \beta
\left( j-1\right) +\beta \left( j+1\right) \right] +\delta _{j0}g_{b}u_{e},
\end{align*}%
where $\delta_{mn}=1$ $(0)$ for $m=n$ $(m\neq n)$. Removing the atomic
amplitudes in the above equation leads to the discrete-scattering equation
of single photons
\begin{subequations}
\label{3cs-03}
\begin{align}
\left( E-\omega _{a}\right) \alpha \left( j\right) & =-\xi _{a}\left[ \alpha
\left( j-1\right) +\alpha \left( j+1\right) \right] \\
& +\delta _{j0}\alpha \left( j\right) V_{a}\left( E\right) +\delta
_{j0}\beta \left( j\right) G\left( E\right),  \notag \\
\left( E-\omega _{b}\right) \beta \left( j\right) & =-\xi _{b}\left[ \beta
\left( j-1\right) +\beta \left( j+1\right) \right] \\
& +\delta _{j0}\alpha \left( j\right) G\left( E\right) +\delta _{j0}\beta
\left( j\right) V_{b}\left( E\right),  \notag
\end{align}
where we have introduced the energy-dependent delta-like potentials $%
V_{d}\left( E\right) \equiv g_{d}^{2}\;V\!\left(E\right) $, with
\end{subequations}
\begin{equation}
V\!\left(E\right) \equiv \frac{\left( E-\omega _{S}\right) }{\left( E-\omega
_{E}\right) \left( E-\omega _{S}\right) -\left\vert \Omega \right\vert ^{2}},
\label{3cs-04}
\end{equation}%
and the effective dispersive coupling strength
\begin{equation}
G\left( E\right) \equiv g_{a}g_{b}V\!\left( E\right)
\end{equation}
between the cavity modes $a_{0}$ and $b_{0}$. It should be pointed out that
the energy of the incident photon indirectly determines whether a repulsive
or attractive potential is localized at $j=0$, as well as the magnitude of
the delta-like potentials and effective coupling strengths. Since $%
V_{d}\left( E\right) $ and $G\left( E\right) $ are induced by the atom, we
rewrite Eq.~(\ref{3cs-04}) as
\begin{equation}
V\left( E\right) =\frac{A_{+}}{E-\omega _{+}}+\frac{A_{-}}{E-\omega _{-}},
\label{3cs-05}
\end{equation}
to capture the effect of the atomic quantum interference, with frequencies $%
\omega _{\pm }=\left( \omega _{S}+\omega _{E}\pm \mu \right) /2$, and
numerators $A_{\pm }=\left[ 1\pm \left( \omega _{E}-\delta_{S}\right) \mu
^{-1}\right] /2$ with
\begin{equation}
\mu =\sqrt{\left( \omega _{E}-\omega _{S}\right) ^{2}+4\left\vert \Omega
\right\vert ^{2}}.  \label{3cs-06}
\end{equation}

\subsection{Dressed states}

Equation~(\ref{3cs-05}) indicates that the classical field dresses the atom
to form doubly-excited states with energies $\omega _{\pm }$ (called dressed
states). The parameter $\mu $ denotes the energy splitting between the two
dressed states. At $E=\omega _{\pm }$, infinite delta potentials are formed
at $j=0$ in both CRWs. It seems that the delta potential would prevent the
propagation of single photons. However, the effective coupling strength $%
G\left( E\right) $ also becomes infinite at $E=\omega _{\pm }$, which may
enable the transfer of the photon from one CRW to the other. When the energy
of the incident photon satisfies the two-photon resonance condition $%
E=\omega _{S}$, both $V_{d}\left( E\right)$ and $G\left( E\right) $ vanish,
the two CRW are decoupled. When the Rabi frequency $\Omega \rightarrow 0$, $%
\omega _{+}\rightarrow \omega _{E}$, and $\omega _{-}\rightarrow \omega _{S}$%
. However, $A_{+}\rightarrow 1$, $A_{-}\rightarrow 0$, i.e., our hybrid
system becomes two CRWs coupled to a two-level system (TLS) in the absence
of the classical field. Infinite delta potentials and an infinite effective
coupling strength between the two CRWs can also be obtained when the energy
of the incident photon is resonant with the TLS. Consequently, it is still
possible for the photon to be transferred from one CRW to the other.
However, it is impossible to decouple the two CRWs.

\subsection{How the router works}

An incident wave impinging upon the left side of one CRW (e.g., $a$) will
result in reflected, transmitted, and transfer waves with the same energy.
The wave functions in the asymptotic regions are given by
\begin{subequations}
\label{3cs-07}
\begin{align}
\alpha \left( j\right) & =\left\{
\begin{array}{c}
e^{ik_{a}j}+r^{a}e^{-ik_{a}j}\text{ \ }j<0 \\
t^{a}e^{ik_{a}j}\text{ \ \ \ \ \ \ \ \ }j>0%
\end{array}%
\right. , \\
\beta \left( j\right) & =\left\{
\begin{array}{c}
t_{l}^{b}e^{-ik_{b}j}\text{ \ }j<0 \\
t_{r}^{b}e^{ik_{b}j}\text{ \ }j>0%
\end{array}%
\right. ,
\end{align}
where $t^{a}$ ($r^{a}$) is the transmitted (reflected) amplitude and $%
t_{l}^{b}$ ($t_{r}^{b}$) is the forward (backward) transfer amplitude. The
relation $E=E_{k}^{[a]}=E_{k}^{[b]}$ between the wavenumbers $k_{a}$ and $%
k_{b}$ can be obtained by applying Eq.~(\ref{3cs-07}) to the discrete
scattering Eq.~(\ref{3cs-03}) far away from the $j=0$ site. However,
applying Eq.~(\ref{3cs-07}) to the discrete scattering Eq.~(\ref{3cs-03}) for
the $0$th and $\pm 1$st sites, we obtain the continuity conditions $%
t_{l}^{b}=t_{r}^{b}\equiv t^{b}$ and $t^{a}=r^{a}+1$, and the scattering
amplitudes
\end{subequations}
\begin{subequations}
\label{3cs-08}
\begin{eqnarray}
t^{a} &=&\frac{2i\xi _{a}\sin k_{a}\left[ 2i\xi _{b}\sin k_{b}-V_{b}\left(
E\right) \right] }{\prod_{d=a,b}\left[ 2i\xi _{d}\sin k_{d}-V_{d}\left( E\right) %
\right] -G^{2}\left( E\right) } \\
t^{b} &=&\frac{G\left( E\right) 2i\xi _{a}\sin k_{a}}{\prod_{d=a,b}\left[ 2i\xi
_{d}\sin k_{d}-V_{d}\left( E\right) \right] -G^{2}\left( E\right) }
\end{eqnarray}
It can be observed that $t^{a}=1$ at $E=\omega _{S}$, which means that the
incident photon with energy $E=\omega _{S}$ will be totally transmitted in
its original CRW due to the vanishing $V_{d}\left( E\right) $ and $G\left(
E\right) $.
\begin{figure}[tbp]
\includegraphics[width=8.5cm]{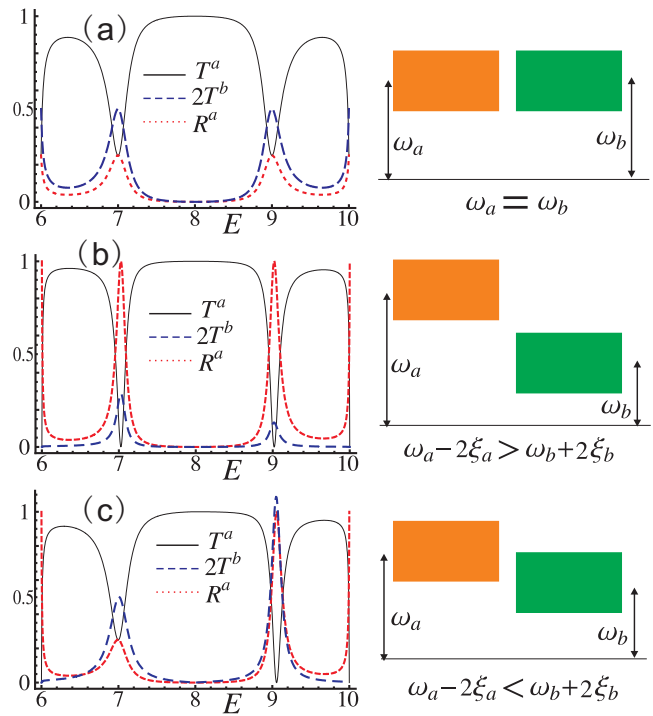}
\caption{(Color online) The transmittance $T^{a}(E)$ (black solid line),
reflectance $R^{a}(E)$ (red dotted line) and total transfer rate $2T^{b}(E)$
(blue dashed line) as a function of the energy of the incident wave. (a) $%
\omega_{b}=8$, the two bands of the bare CRWs are maximally overlapped; (b)
$\omega_{b}=2$, there is no overlap between the two bands; (c) $\omega_{b}=6$,
the two bands are partially overlapped. For convenience, all the parameters
are in units of $\protect\xi_{a}$ and we always set $\xi_{a}=\xi_{b}=1$,
$\omega_{a}=\omega_{E}=\omega_{S}=8$, $\Omega=1$, and $g_{a}=g_{b}=0.5$.}
\label{fig:2}
\end{figure}

In Fig.~\ref{fig:2}, we plotted the transmittance $T^{a}(E)\equiv
|t^{a}(E)|^{2}$, transfer rate $T^{b}\equiv \left\vert t^{b}(E)\right\vert
^{2}$, and reflectance $R^{a}(E)\equiv |r^{a}(E)|^{2}$, as a function of the
incident energy $E$, for three different band configurations of the two bare
CRWs. In Fig.~\ref{fig:2}(a), two bands of the CRWs are maximally
overlapped. It can be found that: (1) the transfer rate does not vanish, and
there is no perfect reflection; therefore single photons incident from the
CRW-$a$ can be transferred to the continuum in the CRW-$b$; (2) Single
photons incident from the left side of the CRW-$a$ can be totally
transmitted to the right side of the CRW-$a$ at $E=\omega_S$. We note that
the total transmission point cannot be observed in the system studied in
Ref.~\cite{zhqrouter}. According to the probability conservation, the
incident flux is required to equal the sum of the reflected, transmitted,
and transfer fluxes. Hence, the scattering amplitudes satisfy $%
|t^a|^2+|r^a|^2+2|t^b|^2=1$. In Fig.~\ref{fig:2}(b), there is no overlap
between two bands. The total transmission still occurs at vanishing
two-photon detuning. However, there are total reflections. It can also be
observed that the flow conservation equation is changed as $%
|t^a|^2+|r^a|^2=1 $, which indicates that the photon flow is confined in the
CRW-$a$. Actually, these total reflections are caused by the incident energy
of a single photon matching the eigenvalues of the bound states of the CRW-$%
b $ with $g_a=0$. These bound states are local modes around the atom in the
CRW-$b$. In Fig.~\ref{fig:2}(c), there is partial overlap between two bands.
The total transmission, total reflection, and transfer to the continuum of
the CRW-$b$ can all be found in Fig.~\ref{fig:2}(c). When the energy of the
incident photon is out of (within) the overlap region of the two continuum
bands, the conservation relation and the related scattering properties are
the same as those in Fig.~\ref{fig:2}(b) [Fig.~\ref{fig:2}(a)].


\section{\label{Sec:4}coherent control of single photons}


In this system, two CRWs provide two 1D continua, each 1D continuum is an
open quantum channel for photons. Without the atom, photons incident from one
quantum channel cannot transfer to the other. In this section, two band
configurations, where energy bands of the CRWs are either maximally
overlapped or have no overlap, are considered separately, to better
understand how the atom controls the flow of photons.
\begin{figure}[tbp]
\includegraphics[width=7.2cm]{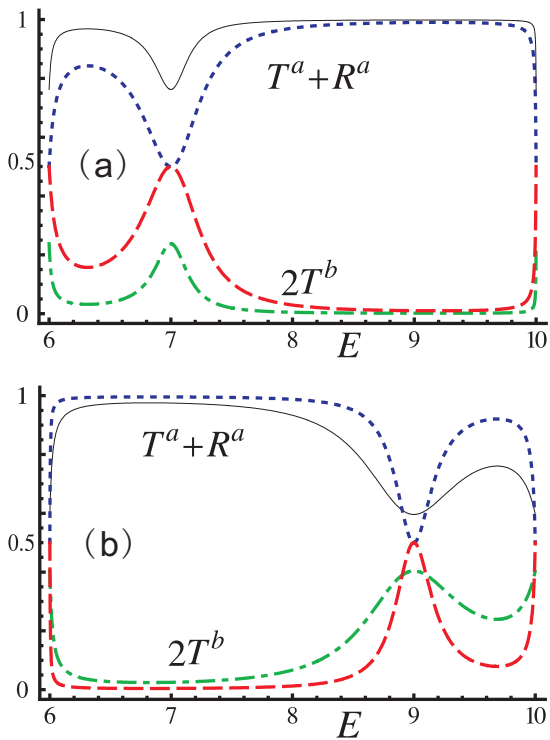}
\caption{(Color online) The coefficients $T^{a}(E)+R^{a}(E)$ (solid black
lines and dotted blue lines), $2T^{b}(E)$ (red dashed lines and
dotted-dashed green lines) as a function of the incident photon energy $E$.
The chosen parameters $\protect\xi_{a}=\protect\xi_{b}=1, \protect\omega_{a}=%
\protect\omega_{b}=8, \protect\omega_{S}=\Omega=0$ indicate that we study
the scattering process for a two-level system interacting with two CRWs,
where the energy band of two CRWs overlap. (a) $\protect\omega_{E}=7$, $%
g_a=0.5$, $g_b=0.5$ $(0.2)$ for the dotted blue and red dashed (solid black
and dotted-dashed green) lines. (b) $\protect\omega_{E}=9$, $g_a=0.4$ $(0.5)$%
, $g_b=0.4$ $(0.8)$ for the dotted blue and red dashed (solid black and
dotted-dashed green) lines.}
\label{fig:3}
\end{figure}

\subsection{Multi-channel quantum router}

We first reveal the underlying physics in Fig.~\ref{fig:2}(a), where two
energy bands are overlapped. Two different situations, where the classical
driving is turned off or on, are considered.
\begin{figure}[tbp]
\includegraphics[width=7.2cm]{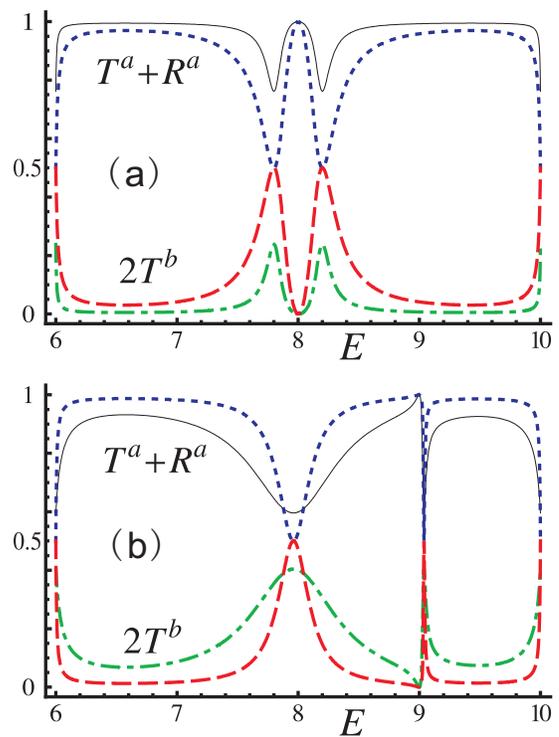}
\caption{(Color online) The coefficients $T^{a}(E)+R^{a}(E)$ (solid black
lines and dotted blue lines), $2T^{b}(E)$ (red dashed lines and
dotted-dashed green lines) as a function of the incident photon energy $E$,
when the classical field is applied. Here, the following parameters $\protect%
\xi_{a}=\protect\xi_{b}=1, \protect\omega_{a}=\protect\omega_{b}=\protect%
\omega_{E}=8, \Omega=0.2$ are fixed. (a) $\protect\omega_S=8$, $g_a=0.5,
g_b=0.2$ $(0.5)$ for solid black and dotted-dashed green (dotted blue and
red dashed) lines. (b) $\protect\omega_S=9$, $g_a=0.5$ $(0.4)$, $g_b=0.8$ $%
(0.4)$ for solid black and dotted-dashed green (dotted blue and red dashed)
lines.}
\label{fig:4}
\end{figure}

When the classical driving is turned off (i.e., $\Omega=0$), single photons
incident from one quantum channel (e.g., the CRW-$a$) will be absorbed by
the atom, which transits from its ground state to its excited state. Since
the excited state is coupled to a continuum of states, the excited TLS will
emit a photon spontaneously into the propagating state of either CRW-$a$ or
CRW-$b$. Consequently, mediated by the atom, photons could be routed from
one quantum channel to the other. In other words, the resonant tunneling
process of the atomic excited state aids the atom to perform quantum
routing. Although it is well-known that the spontaneous emission of the
excited TLS can be exploited to switch the motion of single photons in a
one-dimensional (1D) waveguide~\cite{zhouQdev,FanQdev}, it can also be
exploited to redirect the photons coming from one 1D continuum to the other,
with the TLS mediating the resonant tunneling process. To study this
mechanism, we plot the current flow of the photon in the CRW-$a$ and CRW-$b$
in Fig.~\ref{fig:3}, which are described by the coefficients $%
T^{a}(E)+R^{a}(E)$ and $2T^{b}(E)$, respectively. The nonvanishing transfer
rate around the $E=\omega_E$ shows that when the incident energy $E$
approaches the atomic transition energy $\omega_E$, photons coming from one
CRW are redirected to the other by resonant tunneling. Actually, the atomic
transition energy $\omega_E$ determines the position where the minimum flow
in CRW-a and the maximum of the probability transferred to CRW-b occur in
the energy axis, i.e., the peak of transfer rate is centered at $E=\omega_E$%
. The height of the peak for the transfer rate $2T^{b}(E)$ take the maximal
value when $g_a=g_b$. The width of the peak for the transfer rate $2T^{b}(E)$
is determined by the coupling strengths $g_a$ and $g_b$. The larger the
product $g_ag_b$ is, the wider the peak is. The photonic flow can be nearly
completely confined in the incident CRW once the incident energy of single
photons is largely detuned from the atomic transition energy.

When the classical field is turned on, two dressed states with energies $%
\omega_{\pm}$ are created due to the coupling between a pair of
well-separated atomic bound levels $|e\rangle$, $|s\rangle$ and a classical
field. These dressed states form doubly-excited states of the atom. For an
appropriate Rabi frequency, two dressed states are within the energy bands
of two CRWs. Single photons coming from the CRW-$a$ could excite the atom
from its ground states to either of two dressed states, due to the
transition driven by the cavity mode $a_0$. The spontaneous emission from
the atom provides a chance for photons traveling in both CRWs since the two
dressed states are coupled to the continuums of both CRWs. These tell us
that photons resonantly tunnel from one 1D continuum to the other via two
dressed states, which fulfills the function of quantum routing. The quantum
routing due to resonant tunneling process via the two dressed states could
be observed from the two peaks of the transfer rate in Figs.~\ref{fig:2}(a)
and~\ref{fig:4}. However, different from the case where the classical field
is absent, the photonic flow can be completely confined to the incident CRW,
when $E=\omega_S$, as shown in Fig.~\ref{fig:4}. To know the direction of
the flow in the incident channel, we plot the transmission $T^a(E)$ as a
function of the incident energy in Fig.~\ref{fig:5}. The blue solid (red
dashed) line is the transmission spectra when the classical field is turned
off (on). It can be found that the classical field makes the solid blue line
split into a doublet with a separation of $2\mu$ given in Eq.~(\ref{3cs-06}%
), which is the Autler-Townes splitting~\cite{ATsplit,HaoIanAT}. The
transmission coefficient $T^a(E)$, equal to one at $E=\omega_S$, indicates
that the Autler-Townes splitting yields transparency in a transmission
spectrum. Actually, from the point of view of the ``dressed state''~\cite%
{RMP77(05)00633}, the total transmission appearing at the two-photon
resonance is the result of the interference between the two resonances via
the dressed states.
\begin{figure}[tbp]
\includegraphics[width=7.2cm]{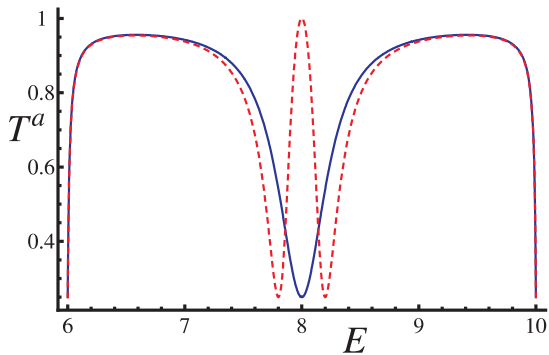}
\caption{(Color online) The transmission coefficient $T^{a}$ in the incident
channel versus the energy $E$ of the incident photon. Here, $\protect\omega%
_S=0$, $\Omega=0$ for the blue solid line, and $\Omega=0.2$ for the red
dashed line. Other parameters are set as follows: $\protect\xi_a=\protect\xi%
_b=1$, $\protect\omega_a=\protect\omega_b=\protect\omega_E=\protect\omega%
_S=8 $, and $g_a=g_b=0.5$. }
\label{fig:5}
\end{figure}

The above discussion tell us that the atom acts as a multi-channel router
for single photons, either in the absence or in the presence of a classical
field, due to the spontaneous emission. However, different from the case
where the classical field is absent, the system with an applied classical
field exhibits quantum interference between two resonances via the dressed
states, which results in the total transmission in the incident channel,
when the energy of single photons satisfy the two-photon resonance. Hence
the classical field can be used to choose the way single photons will take
in this router.


\subsection{Single-channel quantum router}

We now explore the underlying physical mechanism of perfect reflection in
Figs.~\ref{fig:2}(b) and~\ref{fig:2}(c). It can be observed in Eq.(\ref%
{3cs-08}a) that the transmission in the CRW-$a$ vanishes as long as the
condition $2i\xi _{b}\sin k_{b}-g_{b}^{2}V\left( E\right) =0$ is satisfied.
This condition shows that: (1) the parameters related to the CRW-$b$ and the
atom determines the condition for vanishing transmissions, i.e., the
condition is independent of $\omega _{a}$ and $g_{a}$, which characterizes
the CRW-$a$; (2) a real $k_{b}$ cannot meet the condition, and thus the
possibility occurs to the complex extension of the wavenumber $k_{b}$ in the
CRW-$b$. Actually, replacing $k_{b}$ by $(n_{b}\pi+i\kappa _{b})$ yields the
condition
\end{subequations}
\begin{equation}
2\xi_{b}\exp{(in_{b}\pi )}\sinh \kappa _{b}+g_{b}^{2}V\left( E_{\kappa
}\right) =0  \label{4qs-01}
\end{equation}%
for the existence of the bound states in the CRW-$b$, where $\kappa _{b}>0$
and $n_{b}=0,1$. Here $n_{b}=0$ ($n_{b}=1$) indicate that the eigenenergies $%
E_{\kappa }$ lie below (above) the band of the CRW-$b$. Bound states appear
when the translation invariance of the CRW-b is broken. It is the atom
coupled to the CRW-$b$ which breaks down the translation symmetry of the CRW-%
$b$.
\begin{figure}[tbp]
\includegraphics[width=7.2cm]{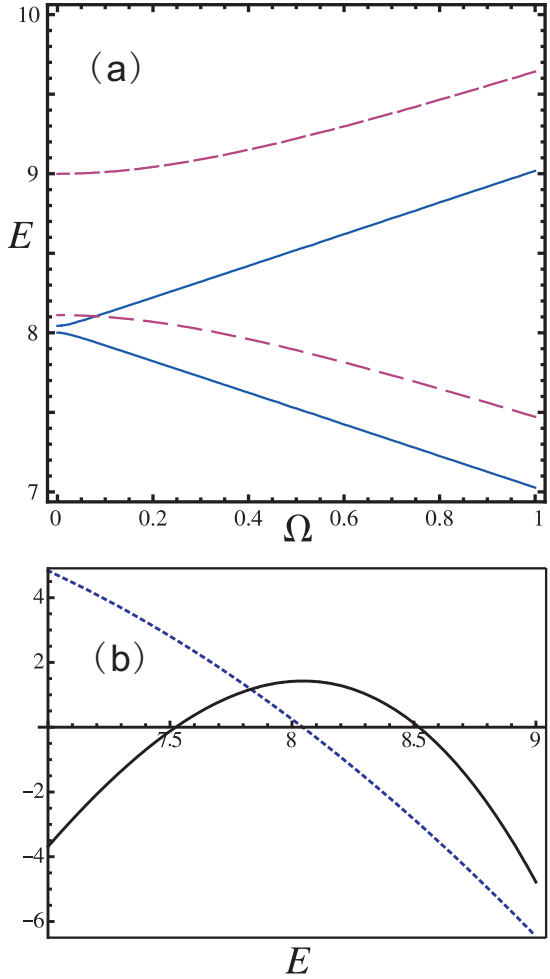}
\caption{(Color online) (a) The solution of Eq.~(\protect\ref{4qs-05}) in
the $(\Omega, E)$ plane. $\protect\omega_S=8, g_a=g_b=0.5$ for the blue
solid curves, $\protect\omega_S=9, g_a=0.5, g_b=0.8$ for the red dashed
curves (b) The left-side of Eq.~(\protect\ref{4qs-05}) ($n_b=1$) as a
function of the energy with $g_a=g_b=0.5$. The points along the transverse
axis indicate the energy of the bound state when $\protect\omega_S=0,
\Omega=0$ for the blue dashed curve and $\protect\omega_S=8, \Omega=0.5$ for
the black solid line. In there figures, we have chosen $\protect\omega_a=8$,
and $\protect\omega_b=2$.}
\label{fig:6}
\end{figure}

To demonstrate that Eq.~(\ref{4qs-01}) is the condition for the existence of
bound states in the CRW-$b$, with a $\Lambda $-system inside, we begin our
study from Eq.~(\ref{3cs-03}b). By setting $g_{a}=0$, we obtain the
discrete-scattering equation for single photons traveling in the CRW-$b$
\begin{eqnarray}
\left( E-\omega _{b}\right) \beta \left( j\right) &=&-\xi _{b}\left[ \beta
\left( j-1\right) +\beta \left( j+1\right) \right]  \label{4qs-02} \\
&&+\delta _{j0}\beta \left( j\right) V_{b}\left( E\right) .  \notag
\end{eqnarray}%
Since the potential $V_{b}\left( E\right) $ vanishes everywhere except at $%
j=0$, the wave function
\begin{equation}
\beta \left( j\right) =\left\{
\begin{array}{c}
D_{1}\exp{\left( j\left( in\pi +\kappa _{b}\right) \right) }\text{ \ for }j<0
\\
D_{2}\exp{\left( j\left( in\pi -\kappa _{b}\right) \right) }\text{ \ for }j>0%
\end{array}%
\right.  \label{4qs-03}
\end{equation}%
must be a damped wave, which decreases exponentially with the distance from
the position $j=0$. Applying the spatial-exponential-decay solution (\ref%
{4qs-03}) to Eq.~(\ref{4qs-02}) far away from the $j=0$ point, we obtain the
dispersion relation%
\begin{equation}
E=\omega _{b}-2\xi _{b}\exp{(in_{b}\pi )}\cosh \kappa _{b}.  \label{4qs-04}
\end{equation}%
And an even parity wavefunction with $D_{1}=D_{2}$ is obtained when applying
Eq.~(\ref{4qs-03}) to the two sites around the zeroth point of Eq.~(\ref%
{4qs-02}). The condition in Eq.~(\ref{4qs-01}) is achieved by inserting the
solution (\ref{4qs-03}) into Eq.~(\ref{4qs-02}) at the $j=0$ point. Using
the dispersion relation in Eq.~(\ref{4qs-04}), the condition for the
existence of the bound states in the CRW-$b$ can be written in terms of the
energy $E_{\kappa }$
\begin{equation}
\left( -1\right) ^{n_{b}}\sqrt{\left( E_{\kappa }-\omega _{b}\right)
^{2}-4\xi _{b}^{2}}+g_{b}^{2}V\left( E_{\kappa }\right) =0.  \label{4qs-05}
\end{equation}
We note that by letting $\Omega=0$, the Eq.~(\ref{4qs-05}) provides the
``existence condition" of the bound states in the CRW-$b$ with an embedded
TLS.

\begin{figure}[tbp]
\includegraphics[width=7.2cm]{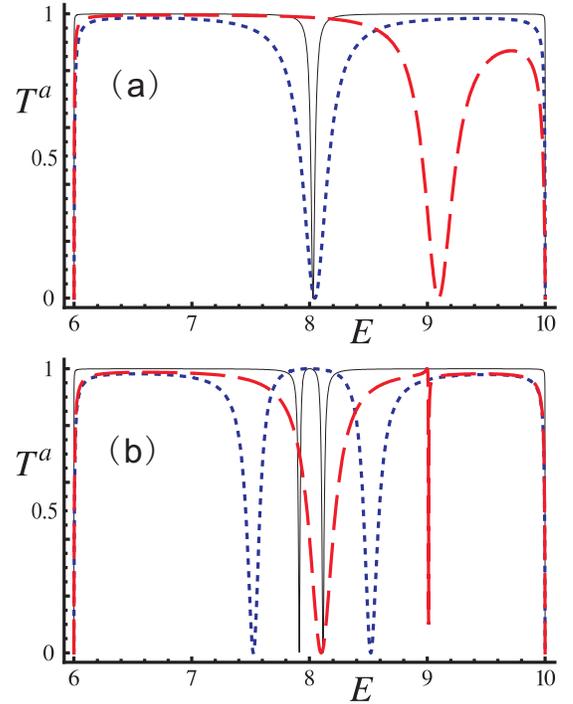}
\caption{(Color online) Transmission coefficient $T^a(E)$ as a function of
the energy $E$ when the classical field is turned off (a) and on (b). In (a)
$\protect\omega_A=8, g_a=0.2, g_b=0.4$ for the black solid line; $\protect%
\omega_A=8, g_a=0.5$, $g_b=0.5$ for the blue dotted line; and $\protect\omega%
_A=9, g_a=0.5, g_b=0.8$ for the red dashed line. (b) $\protect\omega_A=8,
\Omega=0.1, g_a=0.2, g_b=0.4$ for the black solid line; $\protect\omega_A=8,
\Omega=0.5, g_a=0.5, g_b=0.5$ for the blue dotted line; and $\protect\omega%
_A=9, \Omega=0.1, g_a=0.5, g_b=0.8$ for the red dashed line. In all the
figures, we have set $\protect\omega_a=8$, and $\protect\omega_b=2$.}
\label{fig:7}
\end{figure}

In Fig.~\ref{fig:6}(a), we solve Eq.~(\ref{4qs-05}) in the $(\Omega, E)$
plane. It is observed that there are two bound states above the energy band
of the CRW-$b$ for a non-vanishing Rabi frequency. The energy difference
between these two bound states increases as the Rabi frequency increases.
And the larger the one-photon detuning $\Delta=\omega_S-\omega_A$ is, the
wider the energy difference. Figure~\ref{fig:6}(b) present the difference
when the classical field is turn on or off. There is only one bound state
localized around the TLS when $\Omega=0$. It can be found from Fig.~\ref%
{fig:6}(b) that the energies of these two bound states are not the energies $%
\omega_{\pm }$ of the two dressed states. In Fig.~\ref{fig:7}, we plot the
transfer rate $T^b(E)$ as a function of the energy $E$, to show the
difference between the absence and presence of the classical field. It can
be observed that the applied classical field splits the bound state above
the band into two.

In Fig.~\ref{fig:2}(b) and Fig.~\ref{fig:7}, bound states of the CRW-$b$ are
degenerate in energy with the continuum of the CRW-$a$. It is the coupling
between bound states in the CRW-$b$ and the continuum in the CRW-$a$ which
leads to the observation of the bound states via the scattering process. For
single photons incident from the CRW-$a$, the continuum of the CRW-$a$
provides an open channel for the propagation of photons, bound states, on
the other hand, provide close channels. When the energy of the incident
photon matches either of the bound states, the interference between the open
and close channels leads to the total reflection of single photons. We note
that the mechanism of these total reflection is different from the coherent
interference between the incoming wave and the wave emitted by the
doubly-dressed states.

For the band configuration in Fig.~\ref{fig:2}(b), the motion of single
photons from one CRW is confined to the incident CRW. In this case, the atom
functions as a single-photon switch, which routes photons forward or
backward in the incident quantum channel. The scattering process is similar
to the Feshbach resonance in cold atom scattering, where the scattering
cross section diverges when the energy of the incident particle matches the
bound state of the closed channel.

\subsection{Localized photons for the entire system}

Solutions to Eqs.~(\ref{3cs-03}) can be found in the form of either: (i) a
superposition of extended propagating Bloch waves (incident reflected,
transmitted, and transferred by the atom embedded in the CRWs) or (ii)
localized states around the location of the atom. We note that this
localized states are eigenstates of the total system different from the one
obtained above. To show the possibility of the bound state of the total
system, we now consider the case where two bands of the CRWs are not
overlapped. Our purpose here is to derive the condition for the existence of
the bound states. The bound state now is assumed to have the following
solutions with even parity
\begin{align}
\alpha _{\kappa }\left( j\right) & =\left\{
\begin{array}{c}
D\exp{(j \left( in_a\pi +\kappa _{a}\right))}\text{ \ for }j<0 \\
D\exp{(j \left( in_a\pi -\kappa _{a}\right))}\text{ \ for }j>0%
\end{array}%
\right. \\
\beta _{\kappa }\left( j\right) & =\left\{
\begin{array}{c}
C\exp{(j \left( in_b\pi +\kappa _{b}\right))}\text{ \ for }j<0 \\
C\exp{(j \left( in_b\pi -\kappa _{b}\right))}\text{ \ for }j>0%
\end{array}%
\right.
\end{align}
which is localized around the zeroth site, where the TLS is embedded.
Applying the assumed solution to Eq.~(\ref{4qs-02}) far away from the $j=0$
point, we find that $\kappa _a$, and $\kappa _b$ are related to each other
via the energy
\begin{eqnarray}
E_{\kappa}&=&\omega_a-(-1)^{n_a}2\xi_a\cosh\kappa_a  \notag \\
&=&\omega_b-(-1)^{n_b}2\xi_b\cosh\kappa_b
\end{eqnarray}
Applying the assumed solution to Eq.~(\ref{4qs-02}) at the $j=0$ point
yields the final condition for the existence of the bound state in the total
system
\begin{equation}
G^{2}\left( E\right)=\prod_{d}\left[ (-1)^{n_d}2\xi _{d}\sinh
\kappa_{d}+V_{d}\left( E\right) \right],
\end{equation}
which is the denominator of Eq.~(\ref{3cs-08}) with $k_d$ replaced by $%
n_d\pi+i\kappa_d$, where $d=a,b$.

Bound states of the total system provide no contribution to the quantum
transport in the one-quantum subspace, because scattering states survive
only inside the band.

\section{\label{Sec:5}Controllable and scattering-free channels}

In the above discussion, single photons are incident from one quantum
channel. We found that the resonant tunneling process transfers single
photons from one quantum channel to the other when two bands overlap. In
this section, we focus on the overlap band configuration shown in the right
side of Fig.~\ref{fig:2}(a). The purpose now is to investigate the quantum
interference among different quantum channels, and to find the function of
the atom for waves incident from two quantum channels.

We now begin our discussion from the Hamiltonian $H_R$ in the rotating
frame. We first introduce the bright ($B_j$) and dark ($D_j$) modes
\begin{subequations}
\label{4csf-01}
\begin{eqnarray}
B_j&=&a_j\cos\theta+b_j\sin\theta, \\
D_j&=&a_j\sin\theta-b_j\cos\theta,
\end{eqnarray}
which are a linear combination of the cavity-mode operators of both CRWs.
Here, $\tan\theta=g_b/g_a$. In terms of the bright and dark operators and
the condition $\omega_a=\omega_b=\omega, \xi_a=\xi_b=\xi$ for overlap band
configuration, the Hamiltonian of the system reads
\end{subequations}
\begin{eqnarray}
H_R=&&\sum_j\left[\omega B_j^{\dagger}B_j-\xi(B_{j+1}^{\dagger}B_j+h.c.)%
\right]  \notag \\
&&+\sum_j\left[\omega D_j^{\dagger}D_j-\xi(D_{j+1}^{\dagger}D_j+h.c.)\right]
\notag \\
&&+H_A+g\left\vert e\right\rangle \left\langle g\right\vert B_{0} +\Omega
\left\vert e\right\rangle \left\langle s\right\vert +h.c.,  \label{4csf-02}
\end{eqnarray}
where $H_A$ is given in Eq.(\ref{2model-05}) and the coupling strength $%
g\equiv\sqrt{g_a^2+g_b^2}$. Two virtual CRWs (the bright and dark CRWs)
provide the propagating state for single photons. For convenience, the CRW
described by operator $B_j$ $(D_j)$ is called as the ``bright (dark) CRW",
and the quantum channel constructed by the bright (dark) CRW is called as
the bright (dark) channel. The dark CRW is decoupled from the atom.
Consequently, the dark channel is a scattering-free channel, i.e., single
photons incident from the left in the dark CRW are transmitted into the
right with unit probability. However, single photons incident from the
bright CRW could be absorbed by the atom and later emitted spontaneously
into the bright CRW, leading to left-going and right-going photons. This
process is described by a wave with energy $E_k=\omega-2\xi\cos k$, incident
from the left side of the bright CRW, results in a reflected and transmitted
wave in the same CRW.
\begin{figure}[tbp]
\includegraphics[width=7.2cm]{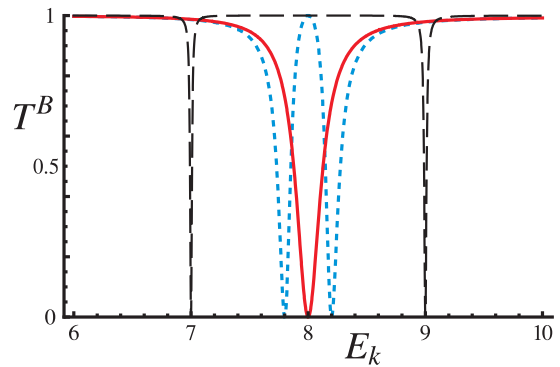}
\caption{(Color online) The transmission coefficient $T^B\equiv|t^B|^2$ as a
function of the energy $E_k$. We set $\protect\xi=1$, $\protect\omega=%
\protect\omega_E=\protect\omega_S=8$, $\Omega=0.2,g=0.5$ for blue dotted
line, $\Omega=0,g=0.5$ for red solid line, and $\Omega=1,g=0.2$ for black
dashed line.}
\label{fig:8}
\end{figure}

By the same approach of Sec.\ref{Sec:3}, the transmission amplitude is
obtained as
\begin{equation}
t^B=\frac{2i\xi\sin k}{2i\xi\sin k -g^2 V(E_k)},
\end{equation}
where $V(E_k)$ is given in Eq.~(\ref{3cs-04}). And the reflection amplitude $%
r_B$ is related to $t_B$ by $t_B=r_B+1$. In Fig.~\ref{fig:8}, we plot the
transmission coefficient as a function of the incident energy. It can be
found that (1) when $\Omega=0$, single photons could be total reflected when
its incident energy is resonant with the atomic transition $%
|e\rangle\leftrightarrow|g\rangle$; (2) When $\Omega\neq 0$, single photons
have a probability $T^B=1$ of being transmitted when the incident energy
satisfies the two-photon resonance $E_k=\omega_S$, and a probability $%
R^B\equiv|r^B|^2=1$ of being reflected when the incident energy matches the
energy $\omega_\pm$ of the dressed states. These total reflections are
caused by the quantum interference between the spontaneous emission from the
atom and the propagating modes in the 1D continuum. The waves emitted by the
doubly-dressed states interfere coherently, such that the back-traveling
wave is eliminated while the forward wave is constructed, which leads to the
perfect transmission of the incident photon. The transmission spectra at $%
\Omega=0$ and $\Omega\neq 0$ indicate that the atom is transparent once the
classical field is applied. Hence, we can control the reflection and
transmission by tuning the Rabi frequency and the classical field frequency
for waves incident from the bright CRW. Hence, we denote the bright channel
as the controllable channel. It should be noted that the propagating states
of the bright channel are orthogonal to those of the dark channel.

\section{\label{Sec:6}discussion and conclusion}

We have studied the coherent scattering process of single photons in two 1D
CRWs. The scattering target is a $\Lambda$-type atom possessing an inversion
center, which fulfill the quantum routing of single photons due to quantum
interference.

When there is an overlap between two bands of the CRWs, the resonant
tunneling process induces the atom to act as a multi-channel router. When
the classical field is absent, one can turn-on the multi-channel routing by
adjusting the transition frequency of the atom between the states $|e\rangle$
and $|g\rangle$ to match the desired propagating states of both CRWs. To
turn off the multi-channel routing, one has to tune the atomic transition
energy far away from the energy of the incident photons. Obviously, the
closure of the multi-channel routing based on large detuning is not perfect.
To perfectly turn off the quantum routing, one can utilize the Autler-Townes
splitting. The procedures are as follows: first, applying a classical field
to drive the transition between the excited state of the atom and an
intermediate level, then adjusting the frequency and the intensity of the
classical field. In addition, the scatter-free and controllable channel are
found for waves incident from both CRWs. Single photons propagate freely in
the scatter-free channel. Without the classical field, the atom acts only as
a perfect mirror for incident waves in the controllable channel due to the
Fano resonance. With the classical field applied, the atom acts not only as
a perfect mirror but also as a transparent medium, i.e., the classical field
selects the photon which is transmitted or reflected.

When there is no overlap between the two bands, the coupling between
discrete energy levels and a continuum makes the atom act as a
single-channel router. When the classical field is absent, the
single-channel router is turned on by adjusting the atomic transition
frequency between states $|e\rangle$ and $|g\rangle$, so that the bound
state in one CRW matches the incident energy of the other CRW. To turn off
the single-channel router, one has to adjust the atomic transition frequency
$\omega_E$ so that the bound states of the CRW are far away from the other
CRW. However, this mechanism for turning-off the single-channel router is
not perfect. With the classical field applied, one cannot only shift the
transmission zeros so that the single-channel router could be operated at
different energies, but also completely turn-off the single-channel router
for single photons with a given energy.

\begin{acknowledgments}
We thank Dr.~S.K.~Ozdemir for his useful feedback on the manuscript.
This work is supported by NSFC No.~11074071, No.~11374095, No.~11105050, and No.~11375060; 
NFRPC No.~2012CB922103, and No.~2013CB921804; 
Hunan Provincial Natural Science Foundation of China (11JJ7001, 12JJ1002),
and Scientific Research Fund of Hunan Provincial Education Department (No.~11B076).
FN is partially supported by the ARO,
RIKEN iTHES Project, MURI Center for Dynamic Magneto-Optics,
JSPS-RFBR contract No.~12-02-92100, Grant-in-Aid for Scientific Research(S),
MEXT Kakenhi on Quantum Cybernetics, and the JSPS via its FIRST program.
\end{acknowledgments}

\end{document}